\documentclass[10pt]{article}

\usepackage{graphicx}
\usepackage{subcaption}

\usepackage[a4paper, left=35mm,right=35mm,top=34mm,bottom=34mm]{geometry}
\usepackage[utf8]{inputenc}
\usepackage[T1]{fontenc}
\usepackage[english]{babel}

\usepackage{enumerate}
\usepackage{graphicx}
\usepackage{hyperref}
\hypersetup{
    colorlinks=true,
    linkcolor=blue,
    filecolor=magenta,      
    urlcolor=cyan,
}

\usepackage{mathtools,amsthm,amssymb,amsfonts}
\usepackage{algorithm}
\usepackage{algpseudocode}
\makeatletter
\def\thm@space@setup{
  \thm@preskip=10pt \thm@postskip=10pt
}
\makeatother
\theoremstyle{plain}

\theoremstyle{plain}

\theoremstyle{definition}

\theoremstyle{definition}

\theoremstyle{remark}

\theoremstyle{remark}

\usepackage{caption} 
\usepackage{booktabs}
\usepackage{listings}
\usepackage{color}
\definecolor{dkgreen}{rgb}{0,0.6,0}
\definecolor{gray}{rgb}{0.5,0.5,0.5}
\definecolor{mauve}{rgb}{0.58,0,0.82}
\usepackage{enumitem}

\newcommand{\email}[1]{\protect\href{mailto:#1}{#1}}

\usepackage{xcolor}[2.11]
\colorlet{inlinkcolor}{green!50!black}
\colorlet{exlinkcolor}{red!50!black}
\if@hidelinks
\hypersetup{ hidelinks = true }
\else
\hypersetup{
  colorlinks = true,
  allcolors = inlinkcolor,
  urlcolor = exlinkcolor,
}
\fi

\newenvironment{@abssec}[1]{
        \vspace{.05in}\parindent .0in
        {\upshape\bfseries #1. }\ignorespaces
    }
    {\par\vspace{.1in}}
\renewenvironment{abstract}{\begin{@abssec}{\abstractname}}{\end{@abssec}}
\newenvironment{keywords}{\begin{@abssec}{Keywords}}{\end{@abssec}}

\usepackage{fancyhdr}

\author{
  {\normalsize Mengchen Wang }\thanks{CentraleSup\'elec, Universit\'e Paris-Saclay, 3 rue Joliot Curie, 91190 Gif-sur-Yvette, France
  (\email{frederic.magoules@hotmail.com})}  \thanks{VENISE Team, LIMSI-CNRS, Universit\'e Paris-Sud, Universit\'e Paris-Saclay, Orsay, France
  (\email{firstname.lastname@limsi.fr})}
  \and
  {\normalsize Nicolas F\'erey\footnotemark[2]}
  \and
  {\normalsize Fr\'ed\'eric Magoul\`es\footnotemark[1]}
  \and
  {\normalsize Patrick Bourdot\footnotemark[2]}
}
\title{Interactive simulation for easy decision-making in fluid dynamics}
\date{}

\begin{document}
\maketitle
\thispagestyle{fancy}

\begin{abstract}
A conventional study of fluid simulation involves different stages including conception, simulation, visualization, and analysis tasks. It is, therefore, necessary to switch between different software and interactive contexts which implies costly data manipulation and increases the time needed for decision making. Our interactive simulation approach was designed to shorten this loop, allowing users to visualize and steer a simulation in progress without waiting for the end of the simulation. The methodology allows the users to control, start, pause, or stop a simulation in progress, to change global physical parameters, to interact with its 3D environment by editing boundary conditions such as walls or obstacles. This approach is made possible by using a methodology such as the Lattice Boltzmann Method (LBM) to achieve interactive time while remaining physically relevant. In this work, we present our platform dedicated to interactive fluid simulation based on LBM. The contribution of our interactive simulation approach to decision making will be evaluated in a study based on a simple but realistic use case.
\end{abstract}


\section{Introduction}

Interactive simulations have been proved to be useful to solve many scientific questions. For instance, \emph{Fold'It}~\cite{foldit2011} helps users to solve the structure of an HIV protein by using an interactive molecular simulation approach. We believe that interactive simulations applied to fluid mechanics can be an interesting tool to support decision making, problem solving and optimization in this field, especially when humans are needed in the loop.

Using Computational Fluid Dynamics (CFD) requires a large set of expertise and tools to design and perform a simulation, then to visualize, explore and analyze the simulation results. Different applications are used sequentially, including 3D modeling software, simulation and visualization software. This process is time-consuming because it involves a lot of technical and unproductive tasks such as data processing and wrapping that require different tools. Moreover, to get accurate relevant physical results, realistic 3D fluid simulations need expansive computational resources. Therefore, it is crucial to detect early errors on the simulation model or the parameters used before the end of the computation. Consequently, it is necessary to regularly extract and analyze simulation snapshots.  By taking advantage of advances in calculation, algorithms, and simulation methodology, an interactive simulation approach was proposed to address these issues, especially to reduce the time lost during this complex process.

There are many works dedicated to interactive data visualization or real-time data visualization in the CFD domain. \emph{ViSTA FlowLib}\cite{10.1145/769953.769963} uses haptic rendering techniques to give a better understanding of the unsteady fluid flows data. Another work~\cite{corsaire2009} provides and evaluates multimodal feedback such as sonification during interaction with fluid simulation, especially to address visual overload.
Stam J. has pioneered the development of the interactive simulation using an unconditionally stable model\cite{10.1145/311535.311548}. Until today, most applications based on interactive simulation approach use particle-based methods such as Translating Eulerian Grids\cite{cohen2010interactive} or Smoothed Particle Hydrodynamics (SPH)\cite{muller2003particle}. These methods target extreme performance, high level rendering, and stability, but at the expense of the physical relevance, which is required by decision-making, physics education or research. Using more relevant methods requires computing centers that are often isolated from the visualization resources which are usually undersized. Even if a lot of technical results were published to support the interactive fluid simulation approach, this methodology remains rarely used. Moreover, there are only a few studies that deal with the problem of the usefulness of interactive simulations in terms of performance and user experience. We propose in this paper a work-in-progress to address this issue by designing an interactive fluid simulation platform based on \emph{Unity 3D} and evaluating the benefit of the interactive simulation approach on decision making from fluid simulation on a simple but realistic use case.

\section{A platform coupling Unity 3D with classical simulation tools}

To provide interactive fluid simulation features, the visualization tool needs to be graphically synchronized with the initial model used as an input to the simulation software. Usually, the user input includes physical properties of the fluid (e.g., speed, pressure, temperature, fluid type), and boundary conditions (e.g., wall, floor, obstacles). During the simulation, the visualization application needs to get the state of the model sent at each time-step from the simulation to update the rendering. To take into account interactive modification on the fluid or its boundary conditions, the interactive visualization tool sends to the simulation the modified fluid states or boundary conditions. According to the rendering and simulation frame rates, it is possible to tune the frequency of data exchange. To realize the communication between two systems, we chose to use a network protocol.

\subsection{A dedicated network protocol and API for interactive fluid simulation}

To be able to integrate any simulation tools daily used by experts, and to ensure the quality of the simulation results, we have developed an API called \emph{CFDriver}\cite{wang2019interactive} embedding a network protocol to perform the data communication between the simulation tool and visualization one.  As a cross-platform API, it works with different programming languages and on different operating systems. It allows an expert to connect our interactive simulation platform to any simulation tools with only a few lines of code. This API provides a distributed architecture to enable the use of computational centers for large fluid simulations. This network protocol allows us to receive the simulation results and also sent the updates from the users to the simulation server during the simulation.

\subsection{Unity 3D as a user interface for interactive simulation}

For the interactive visualization tools, we decided to use \emph{Unity 3D} for our platform. This software was initially designed for game programming but is a good compromise between speed of development and rendering performance. We provide classical camera manipulation tools for rotating, panning, and zooming. For the user's input, we chose to use ray casting to select and change the state of each voxel corresponding to a part of the simulation grid. We use the same cursor to perform multiple tasks such as setting a wall, filling a voxel with water, or emptying a voxel. Besides, it is possible to stop, start or restart, pause or continue the simulation after updating its content such as boundary conditions, grid cell states.

\subsection{Lattice Boltzmann as simulation approach}
We chose to use Lattice Boltzmann Method (LBM) for the simulation tool because it is a good compromise between performance and physical relevance. New solvers such as adaptive relaxation method can simulate more and more conditions like fluid-solid two way coupling simulation\cite{10.1145/3386569.3392400}. The LBM has become very popular for CFD simulations as it can provide fully parallel algorithms\cite{yeomans2006mesoscale}. Since macroscopic variables are translated into particle distribution functions\cite{noble1995consistent}, the simulation is less sensitive to parameter changes during the simulation. Furthermore, this method uses a regular grid which makes the user's input easier to process. Therefore, we can change the parameters cell by cell during the simulation. This method can also provide high performance to achieve interactive rendering times. We use Palabos\cite{latt2020palabos} as our LBM software coupled with Unity 3D.

\section{A user study to compare interactive vs. classical simulations }


The goal of this experiment is to compare the performance between the interactive simulation condition and the non-interactive one. We aimed to measure the added value of interactive simulation in terms of user experience and performance for decision making in a simple water dam design scenario.

\subsection{A water dam design as a use case}

In this scenario, a wall is fixed on the left to stop a water flow coming from the right. In front of the wall, there are different kinds of obstacles to slow down the water wave. Each scene has a certain type of obstacle and different water quantity, corresponding to a different optimal solution in terms of wall height. These scenes allow us to test our hypotheses on different levels of difficulty. The task of the participants was to find the minimum height of the wall to stop the water wave, and optimal solutions were computed by offline simulations performed before the experiment.

\begin{figure}[!h]
  \centering
  \includegraphics[width=1\linewidth]{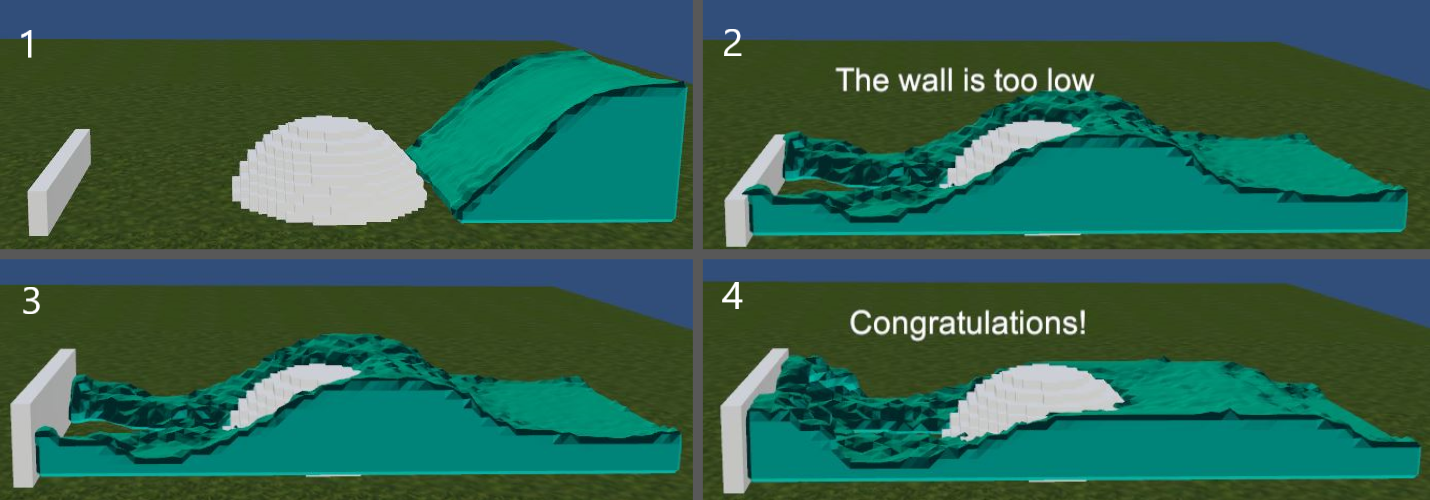}
  \caption{\label{fig:3}
 Interactive simulation of water dam conception: 1 - The simulation starts, 2 - The system detects that water is higher than the wall, 3 - User modifies the wall height, 4 - User finds the optimal height.}
\end{figure}

During the simulation, the system detects if the water overflows the wall or the wall is built too high for the water wave. In this case, the participant has to perform another attempt. Once the system detects that the correct height has been achieved after fluid stabilization, the next scene is loaded (see Fig.\ref{fig:3}). 

In the interactive simulation condition, boundary height can be edited during a simulation in progress to perform the targeted task. In the non-interactive condition, the simulation must be restarted from the initial conditions when modifications are performed. 

The simulation grid is composed of 30 $\times$ 30 $\times$ 72 cells (86400 cells in total), simulating a box whose size is of 1m $\times$ 1m $\times$ 3.2m. The water surface is estimated and rendered using the water fraction method for each voxel/cell of the grid. 

\subsection{Hypothesis}

We expected that the interactive simulation would have better user performance regarding task completion time. We also assumed that the task was easier in the interactive simulation condition. Indeed, this condition allows users to make more attempts with different setups to achieve the task, reducing the time to make a decision, and shortening the loop between simulation, visualization and analysis.

\begin{itemize}
\item H1: Less time is required to find the correct wall height in the interactive simulation condition.
\item H2: Interactive simulation allows users to spend less time to observe and analyze the scene and result.
\item H3: Overall workload is lower with interactive simulation
\item H4: Interactive simulation allows users to perform more attempts to explore larger solution space.
\end{itemize}

\subsection{Experimental process}

Because of the Covid-19 sanitary condition, an experiment in the lab was not permitted. Therefore, before the experiment, an email containing an information notice and a consent form was sent to the subjects. Once they signed the document, they were contacted by phone and invited to connect to our computer remotely via \emph{Teamviewer}. Then, they had to perform a 5-minute familiarisation task with a training scene to familiarize themselves with interactive features and goals. After this training stage, the first session of the experiment started with each mode of simulation (interactive mode or non-interactive condition). After 3 scenes with the same mode of simulation, they were invited to fill the NASA TLX questionnaire\cite{hart2006nasa}. Then the subject entered the second session: another training part with the other mode of simulation started followed by 3 other scenes of the experiment. After the second part, the experiment ended with the same questionnaire for the second session.

\textbf{Participants:} 24 participants (13 males and 11 females) aged from 17 to 36 ($\mu$ = 27, $\sigma$ = 4.83)


\begin{figure}[!h]
  \centering
  \includegraphics[width=1\linewidth]{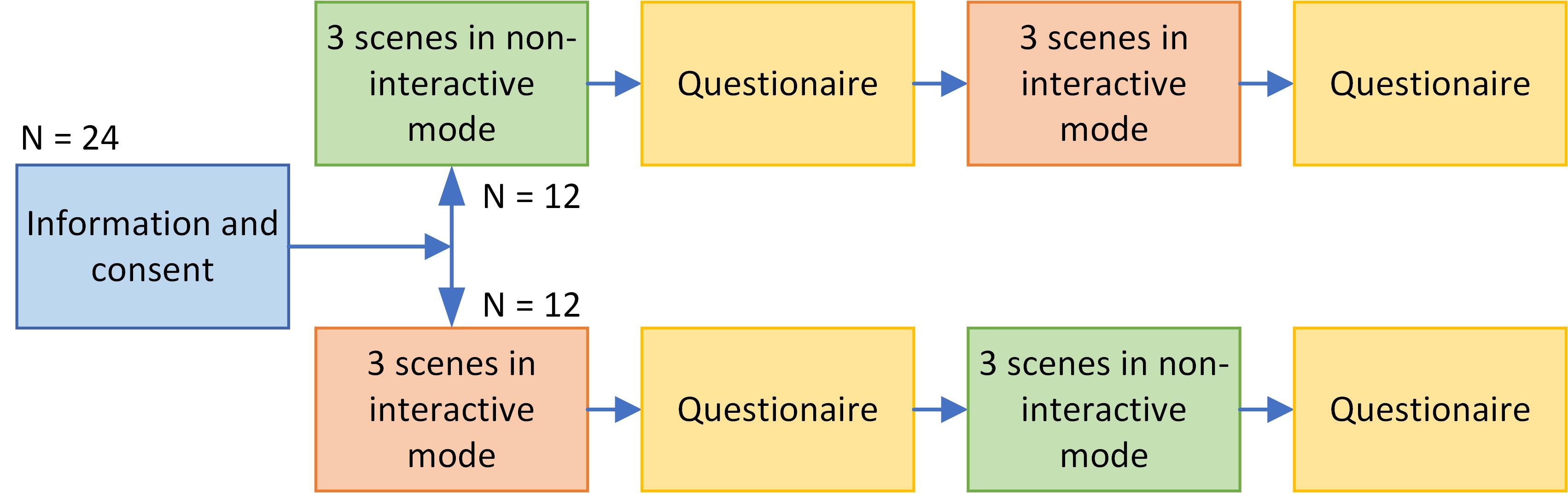}
  \caption{\label{fig:1}
           24 participants were divided in two groups. 12 of them started with interactive simulation then finished with non-interactive simulation. The other 12 started with non-interactive simulation then finished with interactive simulation. The Latin-square design was constructed for counterbalancing the condition and scene.}
\end{figure}

\textbf{Data collection}: For each session of the experiment, we logged all the user activities (changing viewpoint, interaction events, editing boundary condition, etc.), the results received from simulation, and completion time. From this raw data, the following measures were extracted:

\begin{itemize}
\item Task completion time (TCT): The duration for the user to complete the task. The measurement started when the scene was loaded until the user found the optimal height of the wall.
\item Number of failures: The number of failures before achieving the task for each scene.
\item Observation time: The average observation time between each interactive attempt including camera manipulation.
\end{itemize}

\subsection{Results}

The results presented in this section were considered statistically significant when p < 0.05. All the analyses were performed using R. In the bar plots, error bars show the standard deviation (SD).

\textbf{Quantitative results}: We found the data was not all normally distributed in a Shapiro test. In the following part of this section, we performed Wilcoxon Signed Rank test. We registered 144 trials: 2 conditions $\times$ 3 scenes $\times$ 24 participants. 

For TCT, the Wilcoxon Signed Rank test showed that the task was significantly quicker (p < 0.0001) to achieve with interactive simulation (Mean = 230.4, SD = 50.0) than with non-interactive simulation (Mean = 141.625, SD = 23.4).

For the number of failures, the Wilcoxon Signed Rank test showed that the participants tend to try more times (p = 0.0004) with interactive simulation (Mean = 7.58, SD = 1.32) than with non-interactive simulation (Mean = 9.79, SD = 2.41).

For the observation time, the Wilcoxon Signed Rank test showed that the participants spent more time (p = 0.002) with non-interactive simulation (Mean = 8.26, SD = 3.89) than interactive simulation (Mean = 5.66, SD = 3.21).

\textbf{Subjective questionnaire:}
For the NASA-TLX questionnaire, the ratings are averaged to calculate the overall workload. We ran the Wilcoxon Signed Rank test for the overall workload, but we did not found a significant difference (p = 0.346). While the Wilcoxon Signed Rank test showed that participants made less effort (p = 0.02) with interactive simulation (Mean = 3.71, SD = 1.83) than non-interactive simulation (Mean = 4.38, SD = 1.93).

\begin{figure}[!h]
  \centering
  \includegraphics[width=.9\linewidth]{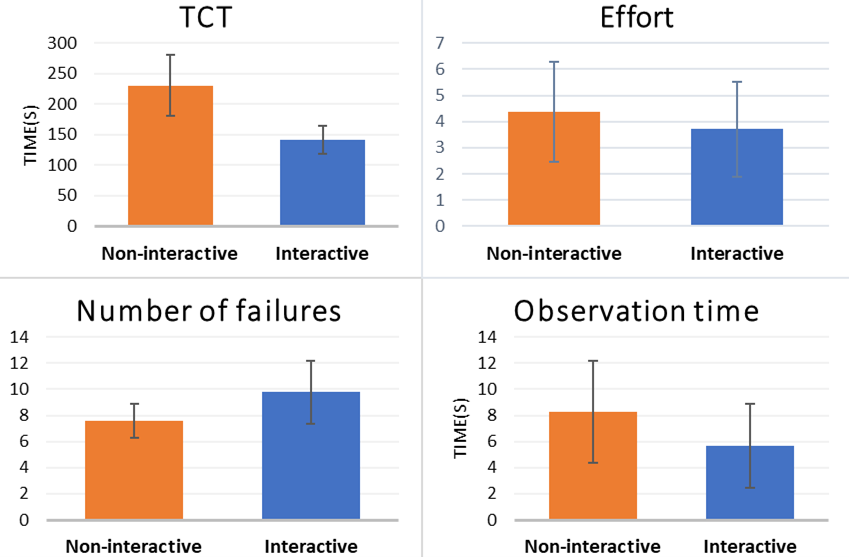}
  \caption{\label{fig:2}
           Mean TCT (top-left), effort (top-right), number of failure (bottom-left) and observation time  (bottom-right). Error bars show standard deviation.}
\end{figure}

\section{Discussion}
From the results, we can first highlight that the participants had 39\% better times completion performances when in the interactive simulation condition (H1). The interactive simulation allows the user to modify the wall height more frequently without restarting the simulation. It is much more efficient for interactive simulation to show the result with the modified wall, as it skipped most parts of the water simulation.

We also found that users made more attempts (H4) and spent less time observing the scene (H2) to achieve the task in the interactive simulation. We interpret this as that the interactive simulation costs less time to make more trials. In non-interactive mode, the participants tried to analyze the result after having the result of the first simulation as they would need to wait longer time to get the results of the modified simulation. Interactive mode let users to make more attempts in limited time to achieve the goal or get a better solution.

On the contrary to our expectations, we did not find significant differences in the overall workload, which rejects H3. From the discussion with the participants, we observed that as the participants were not experts in fluid dynamics, they had difficulties in predicting the movement of the water. We also assumed that the time needed to wait for the end of the simulation in the non-interactive condition might decrease the whole cognitive load. In the meanwhile, we did find that there were significant differences in effort. We interpret this as that since users tended to make more attempts to achieve the task, they did not make efforts to try to understand the simulation. Furthermore, as participants spent less time to achieve the task, it appeared much less difficult to them in performing it.

\section{Conclusion and perspectives}

In this work we present a platform that couples fluid simulation tools with \emph{Unity 3D} to realize the visualization and interaction with a fluid simulation. We address the complex issue of modifying boundary conditions during a simulation in progress by using the LBM. We conducted a first user study to evaluate the added value of interactive simulations compared to the conventional approach in a water dam design scenario. The first results show that users have better performance in achieving the task with an interactive simulation in task completion time. As interactive simulation is less time consuming, more attempts of simulation can replace some work of analyzing the simulation results, which also makes users feel easier to achieve the task. In the meanwhile, the interactive simulation does not reduce the overall users' workload in achieving the task.

Our platform coupling \emph{Unity 3D} and classical simulation tools are presently mainly used to conduct user studies to evaluate the added value of interactive fluid simulation and to compare different contexts such as classical and immersive environments, augmented reality context, or collaborative ones. We plan in short term to conduct further works to confirm the benefit of interactive simulation to larger fluid simulations, including other interaction contexts such as immersive and augmented reality to support collaboration. We will especially test if using ecological movement around the fluid phenomena using tracking and if collaboration features increase the performance and the quality of the decision-making process, and decrease the cognitive load.

This platform also lays the first bricks for a serious game fluid simulation application. Serious games are used for non-entertainment purposes and are applied on different domains \cite{backlund2007games}, such as education, oil and gas application, health care, etc. Serious games enable learners to adapt learning to their cognitive needs and to increase intrinsic motivation of student\cite{malone1981toward}.  \emph{Fold'It} is one of the most famous examples. In our case, our goals are to design scenarios and to work on gameplays to allow students to get a better understanding of the fluid physics phenomena. Inspired by \emph{Fold'It}, we will use our platform as a serious game to illustrate fluid mechanics concepts through appropriate scenarios designed with experts, and to know if serious game and citizen science approaches could solve practical fluid mechanical problems.

\bibliography{ref}
\bibliographystyle{abbrv}

\end{document}